\documentclass[pdflatex,sn-mathphys-num]{sn-jnl}

\usepackage{graphicx}%
\usepackage{multirow}%
\usepackage{amsmath,amssymb,amsfonts}%
\usepackage{amsthm}%
\usepackage{mathrsfs}%
\usepackage[title]{appendix}%
\usepackage{xcolor}%
\usepackage{textcomp}%
\usepackage{manyfoot}%
\usepackage{booktabs}%
\usepackage{algorithm}%
\usepackage{algorithmicx}%
\usepackage{algpseudocode}%
\usepackage{listings}%
\usepackage{appendix}
\usepackage{array}
\usepackage{makecell}
\usepackage{hyperref}

\theoremstyle{thmstyleone}%
\theoremstyle{thmstyletwo}%

\theoremstyle{thmstylethree}%

\begin{document}

\title[Article Title]{Multimodal large language models and physics visual tasks: comparative analysis of performance and costs}


\author{\fnm{Giulia} \sur{Polverini}}
\author{\fnm{Bor} \sur{Gregorcic}\footnote{Corresponding author: \href{mailto:bor.gregorcic@physics.uu.se}{bor.gregorcic@physics.uu.se}}}

\affil{\orgdiv{Department of Physics and Astronomy}, \orgname{Uppsala University}, \orgaddress{\street{Box 516}, \postcode{75120}, \city{Uppsala}, \country{Sweden}}}


\abstract{Multimodal large language models (MLLMs) capable of processing both text and visual inputs are increasingly being explored for uses in physics education, such as tutoring, formative assessment, and grading. This study evaluates a range of publicly available MLLMs on a set of standardized, image-based physics research-based conceptual assessments (concept inventories). We benchmark 15 models from three major providers (Anthropic, Google, and OpenAI) across 102 physics items, focusing on two main questions: (1) How well do these models perform on conceptual physics tasks involving visual representations? and (2) What are the financial costs associated with their use? The results show high variability in both performance and cost. The performance of the tested models ranges from 81.5\%  to as low as 21\%. We also found that expensive models do not always outperform cheaper ones and that, depending on the demands of the context, cheaper models may be sufficiently capable for some tasks. This is especially relevant in contexts where financial resources are limited or for large-scale educational implementation of MLLMs.  By providing these analyses, our aim is to inform teachers, institutions, and other educational stakeholders so that they can make evidence-based decisions about the selection of models for use in AI-supported physics education.}





\maketitle

\newpage




\section{Introduction}\label{Introduction}

\subsection{Large language models in physics education}\label{AI in PE}

Artificial intelligence (AI) has been playing an increasingly prominent role in education. Over the past decade, AI-driven tools have been integrated into a wide range of instructional applications, from intelligent tutoring systems to adaptive learning platforms~\cite{gligorea2023, ahmad2024, liu2025}. AI tools promise greater scalability, personalization, and efficiency—especially in resource-constrained environments~\cite{aderibigbe2023, bozic2023}.

\textit{Large language models} (LLMs), which generate human-like text based on statistical patterns in large-scale datasets~\cite{ejp}, have significantly accelerated this trend. Their improving performance has sparked growing interest in their application to subject-specific education~\cite{zeng2023, xuan2025}—including physics, where researchers have assessed various LLMs capabilities using real-world physics and engineering exams from their educational contexts~\cite{tschisgale2025, yeadon2024impact, dao2023, kortemeyer2023, frenkel2023}.

A growing body of research investigates how LLMs perform across a range of physics education areas. Much of this work has centered on ChatGPT~\cite{openai2025chatgpt}, which has demonstrated notable capabilities across multiple physics-related tasks. Studies examining conceptual reasoning~\cite{gregorcic2024, sirnoorkar2024} and problem solving~\cite{kestin2025, wang2024, kumar2023} have shown that the chatbot can produce coherent and well-structured solutions, although it often still exhibits limitations in replicating human-like sensemaking. In essay-style assignments, ChatGPT-generated responses have reached grading levels comparable to high-performing university students~\cite{yeadon2023}, raising concerns about the validity of take-home examinations. It has also shown strengths in physics-related programming tasks~\cite{kortemeyer2023, yeadon2024comparison}, performing reliably on structured coding tasks, and in lab-based problem solving~\cite{kilde2025, low2023}, analyzing experimental data and carrying out statistical analyses.

Very recently, a new class of models—often referred to as \textit{reasoning language models} (RLMs)~\cite{besta2025} (e.g., OpenAI's o3 and GPT-5, DeepSeek-R1, Alibaba's QwQ)—has been introduced to specifically enhance performance on complex, multi-step tasks. Unlike earlier chatbots, primarily developed for general linguistic fluency, RLMs are designed to imitate step-by-step reasoning, producing intermediate solution steps, applying domain-relevant procedures, and maintaining better consistency throughout the process~\cite{xu2025}. Early evaluations suggest that these models outperform previous generations on benchmark assessments involving mathematical problem solving and STEM-focused conceptual reasoning, leading to a growing interest in their application to physics education~\cite{yoon2025, zhang2025}. However, these capabilities do not come without important limitations, especially in high-complexity problem solving~\cite{illusion-of-thinking}.

Since late 2023, LLMs have also been upgraded to \textit{multimodal large language models} (MLLMs)~\cite{wu2023, wang2024mllm}—systems capable of processing inputs beyond text, such as images, video, and other data. While LLMs are built on a single transformer architecture optimized for sequential text processing, MLLMs integrate separate, modality-specific encoders (e.g., vision transformers for images, transformers for text). These encoders process their respective inputs and output embeddings, which are then aligned and fused—typically via cross-attention or token/feature-level fusion—into a shared representation~\cite{bewersdorff2025}. This allows the model to process inputs (and often also generate outputs) across multiple modalities.

For physics education, the shift toward image processing is particularly relevant. Physics is a discipline that fundamentally relies on a variety of visual representations. Graphs, circuit diagrams, free-body diagrams, vector fields, sketches of experimental setups, and other representations are not merely supplementary, they are integral to physics conceptual reasoning, problem solving, and consequently also play a central role in physics education~\cite{treagust2017}.

The first widely accessible model capable of processing images, ChatGPT-4, attracted early interest from the physics education research community. For instance, in one of the first published evaluations, ChatGPT-4 was tested on the Test of Understanding Graphs in Kinematics (TUG-K)~\cite{tugk}. While the model achieved a performance level comparable to high school students, detailed analyses revealed that its primary reason for failure was the visual misinterpretation of graphs.

Shortly after, with the rise of vision-capable models from a range of AI companies—offered at different prices and performance tiers—comparative evaluations have begun to emerge. These studies examine differences across model families and versions~\cite{bessas2025, jiang2025, perc2024, bema, robledo2024}, languages~\cite{crunchy}, and compare free versus paid access levels~\cite{frontiers}. These comparative studies consistently show that while models can perform well on text-only tasks, their primary bottleneck lies in interpreting visual inputs. In fact, their performance tends to degrade when tasks require image interpretation, regardless of the specific physics subdomain involved~\cite{crunchy}. Instead, accuracy appears to hinge more on the type of  demand posed by the visual representation. For example, Polverini et al.~\cite{bema} highlight that ChatGPT-4o struggles with tasks involving spatial and embodied reasoning (e.g., the use of the right-hand rule).

In addition, a growing number of research studies in physics education have begun to focus on how AI-based systems can support instructors. One of the reasons is that the increasing student/teacher ratios, largely driven by the persistent shortage of qualified instructors~\cite{oecd23}, including in physics~\cite{dewinter2025}, have placed severe pressure on educational institutions and existing instructors. At the same time, there is growing demand for personalized learning: teachers are expected to deliver timely feedback~\cite{paris2022}, tailored guidance~\cite{clark2009}, and fair grading at scale~\cite{tierney2011}, while their capacity is inherently limited~\cite{melnick2008}. Tasks such as grading~\cite{kortemeyer23, kortemeyer2024, mok2024, chen2025}, tips and feedback generation~\cite{chen2024, krupp2024, guo2024}, and support for students with disabilities~\cite{clark2025} often involve analysis of student-drawn representations, generation of explanations coupled to such visual representations, and assessing visually complex responses. MLLMs are increasingly considered as one possible solution to offload teachers from these tasks. 

However, early findings indicate that while MLLMs are capable of spotting common mistakes and offering useful feedback, they often overlook the subtleties of student reasoning and struggle to deliver nuanced, differentiated assessment~\cite{eladawy2024exploring}. Addressing these shortcomings requires carefully designed prompting strategies and other methods aimed at minimizing the impact of their inherent unreliability~\cite{kortemeyer2024assessing}, as well as identifying these models' weaknesses to avoid tasking them with operations that they are ill-suited for. Mok et al.~\cite{mok2024} have found that an MLLM's quality of provided feedback and grading of student solutions of physics problems correlates with the system's own performance on the same problems. In other words, good subject-performance is a prerequisite for educational value. For this reason, it is meaningful to test AIs on tasks that they will be expected to provide feedback on, grade, or tutor students on.

Despite all the recent progress, what we know about how MLLMs perform in different educational situations is still limited and often based on scattered or informal reports. Much of the available information comes from the companies that build these models using proprietary datasets and methods. As a result, it is difficult for educators to judge how well these models would actually work in real educational applications. At the same time, these tools are being widely marketed as ready-to-use solutions for education (e.g.,~\cite{openai2024gpt4o, google2023gemini}), which can garner unrealistic expectations and lead to the use of AI tools in ways that may not be appropriate or effective. To use MLLMs responsibly in teaching and learning, we need independent studies of their performance in different domains of knowledge, including physics.

Alongside exploring the different models' performance, there is also the need to compare the actual cost of use for the tested models. Different MLLMs come with varying pricing, which, while often modest for individual queries, can become significant when scaled to institutional use. However, despite the growing interest in using AI tools in education, there is a lack of studies that systematically evaluate their cost-effectiveness or compare usage-based pricing across models. Understanding these costs is relevant for assessing the feasibility of deploying such models across entire classes or educational programs. Typically, model providers specify usage costs in terms of tokens (i.e., fragments of text or data units), with separate rates for input and output tokens, often expressed as cost per million tokens. However, estimating the number of tokens processed in a given query is not always straightforward~\cite{zhang2024token}. Token count can be influenced not only by the length and complexity of the text but also by the presence of images (in the case of MLLMs) or extended chains of reasoning (in the case of RLMs). These factors make precise cost prediction difficult without detailed usage data, which underscores the need for transparent and accessible pricing analyses for educational planning.

This is relevant because the growing adoption of AI tools in education risks reinforcing existing technological divides~\cite{ahmed2025, unesco2024}. If the most capable models come with prohibitive costs, then only well-funded institutions will be able to afford them, leaving under-resourced schools and students at a disadvantage. This undermines efforts toward equitable access to quality education. By analyzing the performance–cost relationship, we can better understand whether lower-cost models offer sufficient educational value, and help ensure that effective AI-supported learning tools are accessible to a broader range of learners and institutions.

Multimodal AI tools thus display both promise and limitations in physics education. While early evidence suggests that MLLMs may support both teaching and learning physics, questions remain about their reliability and the cost implications of their deployment. These considerations motivate the present study.

\subsection{Research Objectives}\label{Research Objectives}
In this study, we benchmark several publicly available MLLMs—including some that also qualify as RLMs—on a set of conceptual physics questions that involve visual interpretation. By evaluating models from multiple providers under the same conditions, and by pairing those results with the corresponding model usage costs, we aim to equip teachers, institutions, and other stakeholders with the information needed to choose the right tool for their needs and budgets.

We ask the following research questions:

\begin{enumerate}
    \item \textit{How do different MLLMs in mid-2025 perform on conceptual physics tasks that require interpretation of  visual representations?}
    
    Answering this question aims to provide an up-to-date and independent assessment of some of the most widely used MLLMs' capabilities in conceptual physics tasks where visual interpretation is essential. Understanding how the different models perform is critical for physics educators considering their use for tutoring, grading, or feedback generation purposes.
    \\  
    \item \textit{What are the actual financial costs associated with running each of the tested models? }
    
    Answering this question helps determine whether affordable models can offer sufficient performance for educational deployment. Cost is a key consideration for institutions serving large student populations and/or working within tight budgets, and exploring the performance–price trade-off is relevant for sustainable and equitable educational implementation of MLLMs.
\end{enumerate}





\section{Methodology}\label{Methodology}
\subsection{Model selection}\label{Model selection}

To evaluate the performance of MLLMs on image-based physics tasks, we selected a sample of publicly available, vision-capable models from three major providers: Anthropic, Google, and OpenAI. Table~\ref{tab:pricing} summarizes the models included in our study, along with their declared token-based pricing at the time of data collection.

\begin{table*}[htbp]
\small
\caption{Descriptions of the selected AI models. Prices are reported in USD per million tokens.}
\label{tab:pricing}
\begin{tabular*}{\textwidth}{@{\extracolsep{\fill}}p{2cm}p{2.5cm}p{4.3cm}p{1.2cm}p{1.2cm}@{}}
\toprule
\textbf{AI Company} & 
\textbf{Name} & 
\textbf{Model} & 
\textbf{Input Price} & 
\textbf{Output Price} \\
\midrule

\multirow[t]{3}{*}{Anthropic~\cite{anthropic}}
  & Claude Opus 4    & claude-opus-4-20250514     & 15.00  & 75.00 \\
  & Claude Sonnet  4  & claude-sonnet-4-20250514   & 3.00   & 15.00 \\
  & Claude Haiku 3.5 & claude-3-5-haiku-20241022  & 0.80   & 4.00  \\

\midrule

\multirow[t]{5}{*}{Google~\cite{google_gemini,google_gemma}}
  & Gemini 2.5 Pro   & gemini-2.5-pro-preview-06-05   & 1.25  & 10.00 \\
  & Gemini 2.5 Flash & gemini-2.5-flash-preview-05-20 & 0.30  & 2.50  \\
  & Gemini 2.0 Flash & gemini-2.0-flash               & 0.10  & 0.40  \\
  & Gemma 3-27b      & gemma-3-27b-it                 & 0   & 0  \\
  & Gemma 3-4b       & gemma-3-4b-it                  & 0  & 0  \\

\midrule

\multirow[t]{6}{*}{OpenAI~\cite{openai}} 
  & GPT-5              & gpt-5-2025-08-07            & 1.25   & 10.00  \\
  & GPT-5 mini         & gpt-5-mini-2025-08-07            & 0.25   & 2.00  \\
  & GPT-5 nano         & gpt-5-nano-2025-08-07            & 0.05   & 0.40  \\
  & o3              & o3-2025-04-16             & 2.00   & 8.00  \\
  & o4 mini         & o4-mini-2025-04-16        & 1.10  & 4.40  \\
  & GPT-4.1         & gpt-4.1-2025-04-14        & 2.00   & 8.00 \\
  & GPT-4.1 mini    & gpt-4.1-mini-2025-04-14   & 0.40   & 1.60  \\
  & GPT-4.1 nano    & gpt-4.1-nano-2025-04-14   & 0.10  & 0.40  \\
  & GPT-4o          & gpt-4o-2024-11-20         & 2.50   & 10.00 \\

\botrule
\end{tabular*}
\end{table*}

Our selection was guided by the following criteria. First, we included only models that are multimodal and accessible via an application programming interface (API). Second, we aimed to cover a wide performance and pricing spectrum, including both premium-tier systems (e.g., Claude Opus 4, Gemini 2.5 Pro, GPT-5 and o3) and lighter alternatives (e.g., Claude Haiku 3.5, GPT-5 mini, GPT-5 nano, the GPT-4.1 series, and Gemini 2.0 Flash). This also allowed us to explore the cost-effectiveness of different models for potential educational deployment.

Considering potential trade-offs is relevant for accessibility and sustainability reasons. For example, in educational contexts, differences in access to more capable (and expensive) models may contribute to a technological divide, where organizations and individuals with the means to afford more expensive models may be in an advantageous position. Furthermore, if the performance of less expensive and resource-demanding models is as good as that of those that use more resources, it is environmentally responsible to use the less resource-hungry model.

\subsection{Task selection}\label{Task selection}

\begin{figure*}[htb]
\begin{center}
\includegraphics[width=\textwidth]{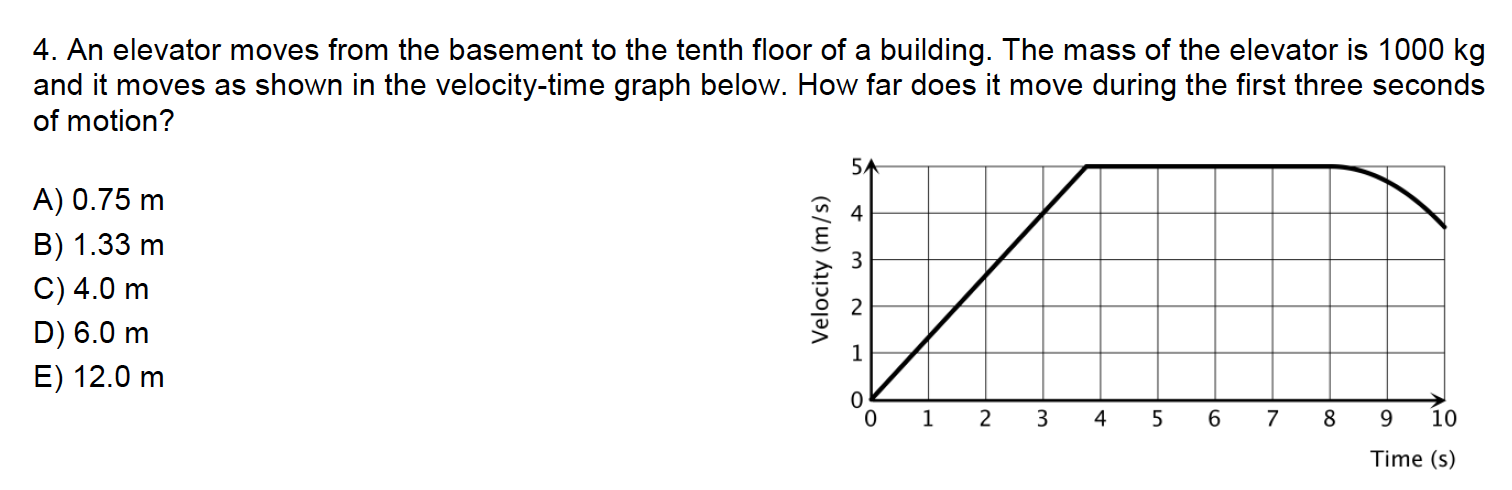}
\end{center}
\caption{An example of one of the 102 test item screenshots submitted to the models. It consists of textual and image parts, as well as a multiple-choice answer list. Image adapted from~\cite{tugk} under the CC-BY 4.0 license. }
\label{fig:TUG}
\end{figure*}

To evaluate model performance on image-based conceptual physics tasks, we selected established concept inventories\footnote{In order to protect the tests' integrity, we do not share them. To
facilitate the interpretation of this paper, we suggest readers access the tests from the \href{https://www.physport.org/}{PhysPort} website.}. Physics concept inventories are research-based multiple-choice assessment tools, developed to probe students’ conceptual understanding of a topic. Each item of the tests typically presents a question and a number of response options. Their standardized structure, focus on conceptual reasoning, often coupled with physics visual representations, makes them well-suited for assessing the capabilities of AI systems in solving visually rich physics tasks. Fig.~\ref{fig:TUG} shows an example of a test item as it was submitted.

For this study, we opted for four tests that cover several areas of undergraduate physics: kinematics (TUG-K), electromagnetism (BEMA), quantum mechanics (QMVI), and geometrical optics (FTGOT). More details about the used inventories are summarized in Table~\ref{tab:pcis}. These tests are well-validated within the physics education research community and widely used across institutions. As such, they represent canonical measures of conceptual understanding, rather than institution-specific assessments. Additionally, they are almost entirely image-based, which aligns with our study’s focus of evaluating the visual processing abilities of MLLMs: 100 out of 102 total items require interpretation of physics visual representations.

\begin{table*}[htbp]
\small
\caption{Presentation of the selected concept inventories, including the number of items and a description of the types of visual representations involved.}
\label{tab:pcis}
\begin{tabular*}{\textwidth}{@{\extracolsep{\fill}}p{2.4cm}p{1.3cm}p{7.5cm}@{}}
\toprule
\textbf{Concept Inventory} & 
\textbf{Number of items} & 
\textbf{Description of visual representations} \\
\midrule

\textbf{BEMA} (Brief Electricity and Magnetism Assessment)~\cite{bema2006}
  & 31 
  & Circuit diagrams, electric field lines, charge distributions, and force vectors. Requires interpretation of symbolic visuals and linking them to field concepts, often in three dimensions. \\ \midrule

\textbf{FTGOT} (Four-Tier Geometrical Optics Test)~\cite{fagotto2017}
  & 20 
  & Ray diagrams and schematic setups involving lenses, mirrors, and light paths. Demands geometrical and spatial reasoning, including estimating distances and angles, tracing ray behavior, and perspective shifts. \\ \midrule

\textbf{QMVI} (Quantum Mechanics Visualization Instrument)~\cite{qmvi2002}
  & 25 
  & Potential energy diagrams, wave functions, probability densities, and energy levels. Requires abstract conceptual mapping between representations. Main focus lies on integrating symbolic and visual representations of quantum phenomena. \\ \midrule

\textbf{TUG-K} (Test of Understanding Graphs in Kinematics)~\cite{tugk1994}
  & 26 
  & Graphs of position, velocity, and acceleration over time. Items typically involve reading or interpreting relationships in line graphs. Involves translating between graphical and linguistic representations of motion, with a focus on slope and area interpretation. \\ \midrule

\botrule
\end{tabular*}
\end{table*}

However, it is important to clarify that this study does not aim to conduct a fine-grained analysis of model performance across the selected physics subdomains. We are not trying to compare, for instance, how well models perform in mechanics versus electromagnetism. This is because our previous observations suggest that performance is not directly linked to the content area itself, but to the type of conceptual reasoning necessary to analyse the visual representations used in a task~\cite{frontiers}. In other words, it is the nature of the visual input—how the model must interpret and reason through it—that plays a more central role than the specific topic.

That being said, there is often an indirect connection: certain types of visual representations tend to be associated with particular areas of physics. See Table~\ref{tab:pcis} for further context: some tests rely heavily on spatial or geometric reasoning, while others emphasize graphical interpretation or symbolic-visual integration. This overlap can give the impression that performance is tied to the subject domain. 

Disentangling this correlation is beyond the scope of our study. Instead, we treat the four test sets as components of a single benchmark that collectively represents a wide range of visual formats used in physics education. While we occasionally report results by individual test, this is done mainly to illustrate that performance differences do exist and may warrant further qualitative analysis. A deeper exploration of model behavior within each subdomain is an important next step—but is not the focus of this work.

\subsection{Data collection}\label{Data Collection}
We captured a screenshot of each item from each test, including the question, the multiple-choice options, and any associated images. For most items, this process was straightforward. However, in the BEMA test a few items shared the same image across multiple questions (16 out of 31). In these cases, we manually separated the items and recreated them so each could stand alone. This adjustment was purely graphical: we simply duplicated the shared image and paired it with each corresponding question. Since none of the items depended on the answers to previous ones, this did not affect the integrity of the questions. Additionally, we edited all FTGOT items to standardize their format. Each original FTGOT item includes four parts: (1) the main question with a multiple-choice list, (2) a confidence rating for that response, (3) a follow-up multiple-choice question asking to explain the initial answer, and (4) another confidence rating related to the explanation. Since our analysis did not focus on reasoning or confidence levels, we excluded parts 2, 3, and 4 from our evaluation. As a result, each FTGOT item was edited to include only the core question, the answer options, and any accompanying images.

Each item of the selected inventories was presented to the models in the form of a screenshot, simulating a visual input scenario aligned with typical student-facing materials.

Every item was submitted 10 times independently in a new context window to each model. This repetition count was chosen based on prior experimentation indicating that MLLMs now mostly exhibit high response consistency across runs. Previous research have demonstrated that models tend to either consistently answer an item correctly or repeatedly select the same incorrect option~\cite{frontiers, bema}. The use of 10 iterations thus balances statistical reliability with cost and environmental sustainability, as earlier protocols involving larger sample sizes are no longer necessary for capturing stable performance patterns.

Inputs were submitted through official APIs. The temperature parameter was set to 0.7 whenever possible to standardize response variability and reflect a moderate level of generative randomness across conditions~\cite{garn2025}. For OpenAI's reasoning models (o3 and o4-mini), the temperature parameter is not modifiable. For GPT-5 series models, the parameter ``reasoning effort" was set to medium. A Python-based script was used to automate the process and record responses into json files~\cite{zenodo}. Answers (selected letter options appearing at the end of the responses) were extracted into csv files for easier processing. This ensured uniformity across trials and streamlined data collection across models. 

To preserve the validity of performance comparisons across models, we deliberately avoided prompt engineering techniques. Each prompt consisted solely of a minimal instruction requesting a clear and structured response:
\\

\textit{Answer the question in the image. If none of the options are correct, answer with the letter N. In a separate last line of the response, restate the answer in the following form: Answer: letter}
\\

The instruction to use the letter \textit{N} when none of the options seem correct was added to reduce artificial inflation of accuracy due to guessing. While students in testing contexts might select an answer regardless of confidence, we sought to focus on the reasoning capabilities of the models rather than the selection of a letter by chance.

This approach was chosen to minimize potential confounding effects introduced by variations in prompt phrasing—a practice that remains largely empirical and difficult to standardize~\cite{ejp}. In particular, we avoided strategies such as few-shot examples~\cite{brown2020}
or explicit Chain-of-Thought (CoT) cues~\cite{wei2022}. As current models often engage in CoT reasoning by default, we focused instead on capturing baseline model behavior.

\subsection{Scoring and analysis}\label{Scoring and Analysis}

Model responses were coded as either correct or incorrect, based on the final selected answer option. We did not apply any conditional grading, even though it was suggested in a small number of items on one of the four assessments (i.e., BEMA). Answers containing the letter \textit{N} were considered incorrect.

Thanks to the consistent output format requested in the prompt, final answers could be efficiently parsed using the last line of the model’s response (e.g., “Answer: B”). However, minor post-processing was required in a small number of cases where the model restated the full text of the chosen answer instead of the corresponding letter. These responses were manually reviewed and coded according to the appropriate option. In no cases did models return multiple selections or unrelated text.

To verify response quality, both authors independently reviewed a randomly selected subset (approximately 30\%) of all responses. These checks confirmed that the models consistently followed the requested response format and that no systematic issues were present.

We computed the performance of each model on each concept inventory by first calculating the percentage of correct responses for each item (based on 10 repeated runs), and then averaging these per-item scores across the full inventory. Each concept inventory contained a fixed number of items (102 in total across all tests), and all models completed all items without submission failures. This item-level averaging approach ensures that each question contributes equally to the final score, regardless of individual item difficulty or model consistency. In addition to the average accuracy, we computed the standard deviation (SD) and standard error of the mean (SEM) for each model on each test. These were calculated from the 10-run distribution for each item and then aggregated across items, providing a measure of response variability and confidence in the estimated performance levels. All results are reported as raw percentages, without normalization or scaling.

Importantly, this study does not analyze the reasoning or explanations provided in the model outputs. While responses often included extended justifications, our scoring procedure was based solely on the final selected answer. This means we did not assess whether the model’s reasoning was scientifically accurate, coherent, or aligned with its answer. As such, it is entirely possible that a model selected the correct option for incorrect reasons—or vice versa. This design choice reflects the quantitative nature of our analysis, which aims to establish baseline performance levels and relate them to practical considerations such as model cost and accessibility. A full qualitative analysis of reasoning accuracy, coherence, or potential biases would require a different methodological framework and falls outside the scope of this study. For the same reason, we do not offer detailed interpretations of why models performed differently across the four concept inventories. While we propose potential explanations in light of previous research, these remain speculative and are not derived directly from our results.

We determined the cost for running each model by tracking the number of input and output tokens for each model and multiplying them by the listed per-token prices. In reporting the model costs, we normalized the cost to represent the expected cost of running our benchmark on the model a single time—i.e., submitting all 102 items once and receiving one response for each item. The costs reported in this paper include both input and output costs. Calculating the expected input costs was straightforward; we simply multiplied the number of tokens for each input (which was the same for all 10 iterations of an item on a given model) with the model-specific input price per token (see Table~\ref{tab:pricing} for token-based prices). For calculating the output costs, we took the average number of tokens outputted by a model for a given item across all 10 iterations, and summed these item averages across all items for each model to obtain the expected number of tokens for the entire benchmark. Multiplying this number with the price per output token for the model gave us the expected output cost of running the entire benchmark for that model. For reasoning models, reasoning tokens are counted and charged as output tokens, even though they are not directly accessible to the user. In these models, reasoning tokens represented most of the cost.




\section{Results}\label{Results}
\subsection{Performance analysis}\label{Performance analysis}
The results in Table~\ref{tab:data} show performance outcomes for a set of MLLMs tested across four physics concept inventories. For each model, the table reports percentage accuracy (Perc), along with standard deviation (SD) and standard error of the mean (SEM) based on 10 repeated runs per item. Each item's performance was treated as an independent Bernoulli variable with an experimentally determined success probability (item score). SD was calculated by taking the square root of the sum of item score variances, and SEM was calculated by dividing the SD by the square root of 10 (number of repetitions of each item). Note that the SD of the performance of each model is a consequence of the variability of the model output and is not expected to decrease with further increasing number of repetitions of each item. The SEM, on the other hand, would further decrease.

Overall, there is substantial variation in performance across models. GPT-5 was the best performing model (81.5\%). o3, Gemini 2.5 Pro and GPT-5 mini yielded high average scores (76.2\%, 75.8\%, and 75.0\% respectively), followed closely by o4 mini (71.5\%). Models such as Gemini 2.5 Flash (66.8\%), GPT-5 nano (60.7\%), Claude Opus 4 (57.0\%), GPT-4.1 mini (53.8\%), Claude Sonnet 4 (53.7\%), and GPT-4.1 (52.5\%) fall in the middle range. Others, like Claude Haiku 3.5 (28.2\%), GPT-4.1 nano (25.0\%), and Gemma 3-4b (21.0\%) scored considerably lower. The range between the highest and lowest average model score exceeds 60 percentage points, highlighting the breadth of performance observed across the models.

\clearpage
{
\pagestyle{empty}

\begin{table*}[htbp]
\small
\caption{Percentage performance (Perf), standard deviation (SD), and standard error of the mean (SEM) for each model on the selected concept inventories. Models are ordered by decreasing total score (Tot AI). The last row represents the average performance of all MLLMs on each concept inventory constituting the total benchmark.}
\label{tab:data}
\begin{tabular*}{\textwidth}{@{\extracolsep{\fill}}p{3.2cm}p{1.2cm}cccccc@{}}
\toprule
\textbf{Model} & 
 & 
\textbf{BEMA} & 
\textbf{TUG-K} & 
\textbf{QMVI} & 
\textbf{FTGOT} & 
\textbf{Tot AI} \\
\midrule

\multirow[t]{3}{*}{GPT-5}
  & \textbf{Perf} & \textbf{93.2} & \textbf{92.3} & \textbf{82.0} & \textbf{48.5} & \textbf{81.5} \\
  & SD   & 3.4  & 3.7  & 4.8  & 7.4  & 2.3 \\
  & SEM  & 1.1  & 1.2  & 1.5  & 2.3  & 0.7 \\[1ex]
  
  \multirow[t]{3}{*}{o3}
  & \textbf{Perf} & \textbf{87.4} & \textbf{88.5} & \textbf{74.0} & \textbf{45.5} & \textbf{76.2} \\
  & SD   & 4.4  & 3.9  & 6.0  & 6.5  & 2.6 \\
  & SEM  & 1.4  & 1.2  & 1.9  & 2.1  & 0.8 \\[1ex]

\multirow[t]{3}{*}{Gemini 2.5 Pro}
  & \textbf{Perf} & \textbf{87.4} & \textbf{84.2} & \textbf{72.0} & \textbf{51.5} & \textbf{75.8} \\
  & SD   & 3.8  & 5.1  & 5.8  & 7.0  & 2.6 \\
  & SEM  & 1.2  & 1.6  & 1.8  & 2.2  & 0.8 \\[1ex]

  \multirow[t]{3}{*}{GPT-5 mini}
  & \textbf{Perf} & \textbf{91.6} & \textbf{81.5} & \textbf{77.2} & \textbf{38.0} & \textbf{75.0} \\
  & SD   & 4.4  & 5.8  & 4.0  & 6.2  & 2.6 \\
  & SEM  & 1.4  & 1.8  & 1.2  & 2.0  & 0.8 \\[1ex]

\multirow[t]{3}{*}{o4 mini}
  & \textbf{Perf} & \textbf{91.9} & \textbf{73.5} & \textbf{69.6} & \textbf{39.5} & \textbf{71.5} \\
  & SD   & 4.0  & 4.6  & 4.2  & 7.1  & 2.4 \\
  & SEM  & 1.3  & 1.4  & 1.3  & 2.3  & 0.8 \\[1ex]

\multirow[t]{3}{*}{Gemini 2.5 Flash}
  & \textbf{Perf} & \textbf{77.7} & \textbf{80.8} & \textbf{60.4} & \textbf{39.5} & \textbf{66.8} \\
  & SD   & 4.7  & 6.1  & 7.0  & 5.7  & 3.0 \\
  & SEM  & 1.5  & 1.9  & 2.2  & 1.8  & 0.9 \\[1ex]

\multirow[c]{3}{*}{\shortstack[l]{Gemini 2.5 Flash\\(no reasoning)}}
  & \textbf{Perf} & \textbf{75.5} & \textbf{75.8} & \textbf{52.0} & \textbf{38.5} & \textbf{62.5} \\
  & SD   & 4.4  & 6.5  & 6.8  & 7.5  & 3.1 \\
  & SEM  & 1.4  & 2.1  & 2.1  & 2.4  & 1.0 \\[1ex]

\multirow[t]{3}{*}{GPT-5 nano}
  & \textbf{Perf} & \textbf{77.4} & \textbf{71.2} & \textbf{53.2} & \textbf{30.5} & \textbf{60.7} \\
  & SD   & 4.9  & 5.8  & 6.4  & 8.6  & 3.1 \\
  & SEM  & 1.5  & 1.8  & 2.0  & 2.7  & 1.0 \\[1ex]

\multirow[t]{3}{*}{Claude Opus 4}
  & \textbf{Perf} & \textbf{70.0} & \textbf{68.8} & \textbf{43.2} & \textbf{38.5} & \textbf{57.0} \\
  & SD   & 4.9  & 5.1  & 5.9  & 7.9  & 2.9 \\
  & SEM  & 1.5  & 1.6  & 1.9  & 2.5  & 0.9 \\[1ex]

\multirow[t]{3}{*}{GPT-4.1 mini}
  & \textbf{Perf} & \textbf{64.5} & \textbf{68.8} & \textbf{41.6} & \textbf{33.0} & \textbf{53.8} \\
  & SD   & 4.8  & 6.9  & 6.9  & 8.0  & 3.2 \\
  & SEM  & 1.5  & 2.2  & 2.2  & 2.5  & 1.0 \\[1ex]

\multirow[t]{3}{*}{Claude Sonnet 4}
  & \textbf{Perf} & \textbf{66.8} & \textbf{66.5} & \textbf{37.6} & \textbf{37.0} & \textbf{53.7} \\
  & SD   & 4.2  & 6.0  & 6.7  & 7.0  & 2.9 \\
  & SEM  & 1.3  & 1.9  & 2.1  & 2.2  & 0.9 \\[1ex]

\multirow[t]{3}{*}{GPT-4.1}
  & \textbf{Perf} & \textbf{60.3} & \textbf{73.1} & \textbf{32.4} & \textbf{38.5} & \textbf{52.5} \\
  & SD   & 4.2  & 5.5  & 6.5  & 5.0  & 2.7 \\
  & SEM  & 1.3  & 1.7  & 2.1  & 1.6  & 0.8 \\[1ex]

\multirow[t]{3}{*}{Gemini 2.0 Flash}
  & \textbf{Perf} & \textbf{61.3} & \textbf{55.8} & \textbf{34.4} & \textbf{35.5} & \textbf{48.2} \\
  & SD   & 4.6  & 6.4  & 6.3  & 6.9  & 3.0 \\
  & SEM  & 1.4  & 2.0  & 2.0  & 2.2  & 0.9 \\[1ex]

\multirow[t]{3}{*}{GPT-4o}
  & \textbf{Perf} & \textbf{57.1} & \textbf{50.4} & \textbf{22.4} & \textbf{36.5} & \textbf{42.8} \\
  & SD   & 5.0  & 5.6  & 6.4  & 7.5  & 3.0 \\
  & SEM  & 1.6  & 1.8  & 2.0  & 2.4  & 0.9 \\[1ex]

\multirow[t]{3}{*}{Gemma 3-27b}
  & \textbf{Perf} & \textbf{46.1} & \textbf{59.2} & \textbf{14.8} & \textbf{14.0} & \textbf{35.5} \\
  & SD   & 4.8  & 5.5  & 3.7  & 4.6  & 2.4 \\
  & SEM  & 1.5  & 1.7  & 1.2  & 1.5  & 0.8 \\[1ex]

\multirow[t]{3}{*}{Claude Haiku 3.5}
  & \textbf{Perf} & \textbf{40.3} & \textbf{31.2} & \textbf{20.8} & \textbf{15.0} & \textbf{28.2} \\
  & SD   & 6.0  & 5.4  & 5.8  & 5.3  & 2.9 \\
  & SEM  & 1.9  & 1.7  & 1.8  & 1.7  & 0.9 \\[1ex]

\multirow[t]{3}{*}{GPT-4.1 nano}
  & \textbf{Perf} & \textbf{31.6} & \textbf{29.2} & \textbf{8.0}  & \textbf{30.5} & \textbf{25.0} \\
  & SD   & 6.0  & 7.5  & 5.2  & 7.6  & 3.3 \\
  & SEM  & 1.9  & 2.4  & 1.6  & 2.4  & 1.0 \\[1ex]

\multirow[t]{3}{*}{Gemma 3-4b}
  & \textbf{Perf} & \textbf{23.2} & \textbf{16.9} & \textbf{19.2} & \textbf{25.0} & \textbf{21.0} \\
  & SD   & 3.5  & 2.7  & 3.2  & 0.0  & 1.5 \\
  & SEM  & 1.1  & 0.9  & 1.0  & 0.0  & 0.5 \\

\midrule
\multirow[c]{3}{*}{\shortstack[l]{Avg AI\\(per concept inventory)}}
  & \textbf{Perf} & \textbf{66.7} & \textbf{64.4} & \textbf{44.4} & \textbf{35.1} & -- \\
  & SD   & 4.6  & 5.5  & 5.7  & 6.3  & -- \\
  & SEM  & 1.4  & 1.7  & 1.8  & 2.0  & -- \\

\botrule
\end{tabular*}
\end{table*}

\clearpage
}

\begin{figure*}[htb]
\begin{center}
\includegraphics[width=\textwidth]{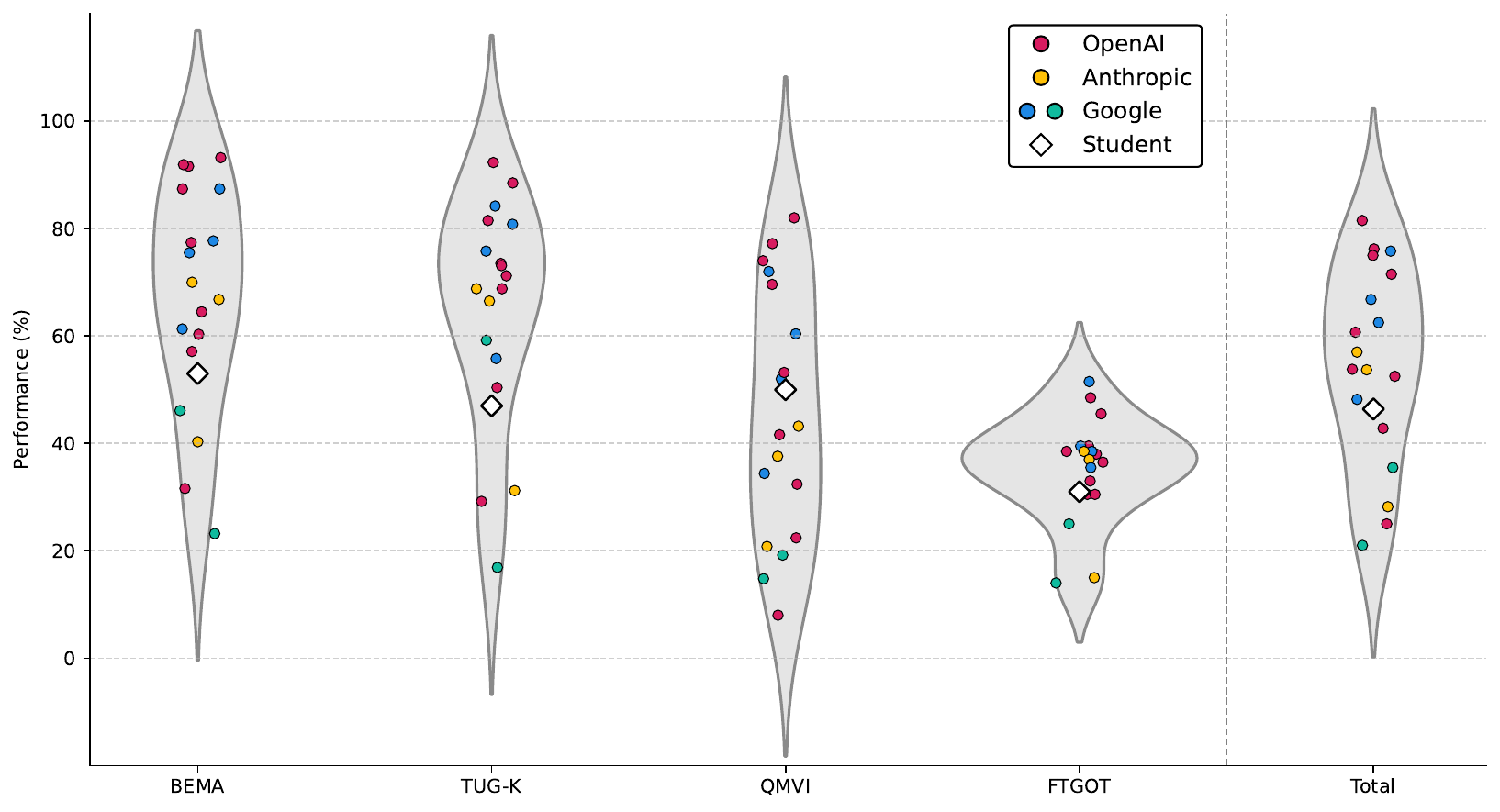}
\end{center}
\caption{Distribution of model scores across the selected inventories, together with the average distribution (right-most violin). Individual models are color-coded according to their developer. The full set of data is available in Table~\ref{tab:data}. For reference, post-instruction university student performance is marked with a white diamond. Student data from~\cite{wheatley2024, zavala2017, qmvi2002, fagotto2017}.}
\label{fig:violin}
\end{figure*}

Performance also varied across concept inventories (Fig.~\ref{fig:violin}). The average performance across all tested models on BEMA (66.7\%) and TUG-K (64.4\%) was higher than on QMVI (44.4\%) and FTGOT (35.1\%). These patterns are consistent across most models. For example, GPT-5 performed above 90\% for BEMA and TUG-K, about 10 percentage points lower on QMVI, and less than 50\% on FTGOT.  Following a similar pattern, o3 and Gemini 2.5 Pro exceeded 84\% on both BEMA and TUG-K, while none of them surpassed 75\% on QMVI or 52\% on FTGOT. Across inventories, the maximum difference between a given model’s highest and lowest test score ranged from 30 to 50 percentage points, suggesting some inventories (especially FTGOT) have higher difficulty even for high-performing models.

The statistical measures provide additional detail. SDs were mostly between 3 and 7 percentage points of the total available points, with most SEM values below 2.5\%. In many cases, higher-scoring models exhibited both higher average performance and lower spread of scores, indicating greater consistency across runs. Some lower-performing models also showed relatively low variability of scores, suggesting consistent but inaccurate responses. In contrast, several mid-range models displayed greater variability, reflecting a mix of correct and incorrect answers across trials.

Looking at performance within model families, some patterns emerge. Among the OpenAI models, GPT-5, o3, GPT-5 mini, o4 mini and even GPT-5 nano, performed better than GPT-4.1 and GPT-4.1 mini, while GPT-4.1 nano scored lowest. Similarly, for Google's models, Gemini 2.5 Pro outperformed Gemini 2.5 Flash, and Gemini 2.5 Flash outperformed the earlier Gemini 2.0 Flash, as well as Gemma. For Anthropic, Claude Opus 4 performed better than Claude Sonnet 4, while Claude Haiku 3.5 scored lowest in that family. While exact differences vary, in most cases each successive tier within a provider correlates with a performance increase of 10–20 percentage points.

\subsection{Cost analysis}\label{Cost analysis}

\begin{figure*}[htb]
\begin{center}
\includegraphics[width=\textwidth]{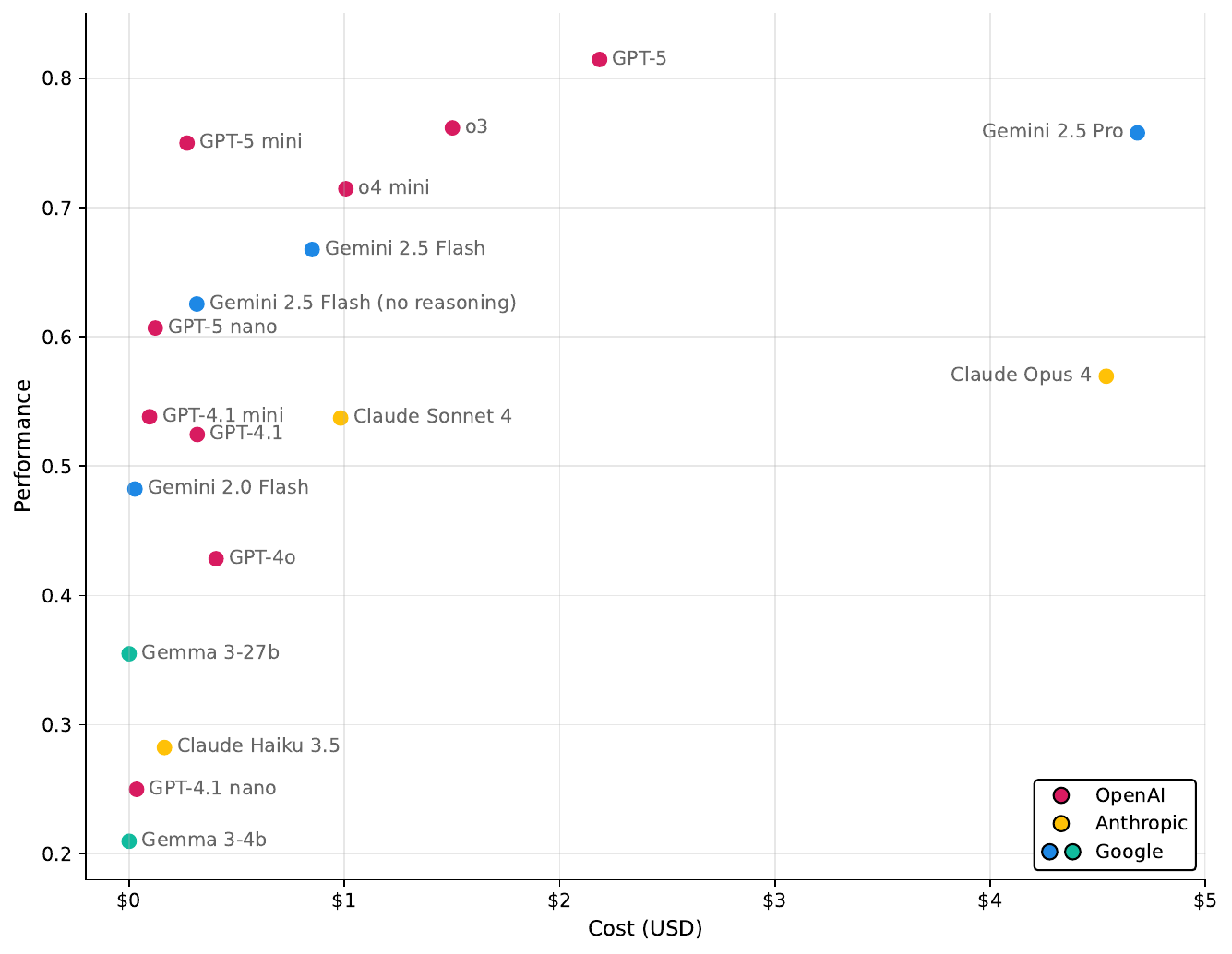}
\end{center}
\caption{Tested MLLMs' distribution of their total performance on the benchmark (vertical axis) and the average cost of running the benchmark once for each model (horizontal axis).}
\label{fig:scatter}
\end{figure*}

To better understand the relationship between model performance and usage cost, we plotted each model's average score against its cost. The resulting plot in Fig.~\ref{fig:scatter} reveals a few patterns. For a more detailed breakdown of the costs, see Table~\ref{tab:token-cost}.

First, while there is an overall correlation between cost and performance, the relationship is not proportional, and there are some interesting outliers. GPT-5 is the best performing model with an accuracy score of 81.5\% at the cost of \$2.18. Some of the other relatively well-performing models, such as o3 and o4 mini, are positioned in the cost range between \$1.00 and \$1.50. Despite not being the most expensive, they achieved accuracy scores of 76.2\% and 71.5\%, respectively, which is comparable to the much more costly Gemini 2.5 Pro model (\$4.68), with a performance of 75.8\%. The clear outlier here is GPT-5 mini, with the accuracy score of 75\% at an outstandingly low cost of \$0.26. On the other hand, Claude Opus 4, only marginally cheaper than Gemini 2.5 Pro, achieved only 57.0\% on the benchmark, significantly underperforming relative to its cost.

Second, a cluster of low-cost, low-performing models is also evident. For instance, the Gemma models, which are freely available (no cost to the user), occupy the bottom of the performance scale (21.0–35.5\%). Although they may be valuable in other contexts, they are currently not competitive for physics conceptual tasks requiring visual interpretation. 

Third, several models strike a compelling balance between affordability and capability. Gemini 2.5 Flash, when the ``reasoning is enabled", performs in the high 60\% range while maintaining the cost just below \$1.00. Even more interesting is the performance of Gemini 2.5 Flash with ``reasoning disabled". It achieved a performance in the lower 60\% range at the cost of only \$0.31, making it an interesting contender for low-cost uses. For less demanding uses or tight budgets, GPT-5 nano appears to be a good candidate, with low cost and performance just above 60\%. Note that this is 38 times cheaper than Claude Opus 4, which performs at only 57\%. Interestingly, GPT-4.1 mini outperformed its bigger brother, GPT-4.1, and displayed performance in the mid-50\% range at the price of only \$0.10. Once again, it is notable that GPT-5 mini is in a class of its own at this price point, performing about 10 percentage points better than other models at the cost of \$0.27. 

Finally, grouping by provider reveals broader trends. OpenAI's models tend to cluster in the upper-left quadrant—combining solid performance with reasonable costs. Google's lineup is more varied, with models ranging from the very strong but expensive Gemini 2.5 Pro, to the underperforming but completely free Gemma models. Anthropic's Claude models fall in the middle or lower performance tiers, despite being relatively expensive, suggesting a less favorable price-to-performance ratio for this particular task set.

\begin{table*}[htbp]
\small
\caption{Token usage and cost by model. Token usage gives the average number of tokens (input, output, and reasoning, where applicable) for one iteration of the 102 items in the benchmark test.  Costs are given in USD. The final column shows the total cost of running all the 102 items once for each model.}
\label{tab:token-cost}
\begin{tabular*}{\textwidth}{@{\extracolsep{\fill}}%
    >{\raggedright\arraybackslash}m{2.5cm}  
    >{\raggedright\arraybackslash}m{1cm}    
    >{\raggedright\arraybackslash}m{1cm}    
    >{\raggedright\arraybackslash}m{1cm}    
    >{\raggedright\arraybackslash}m{1cm}    
    >{\raggedright\arraybackslash}m{1cm}    
    >{\raggedright\arraybackslash}m{1cm}    
    >{\raggedright\arraybackslash}m{1cm}    
@{}}
\toprule
\makecell{\textbf{Model}} & 
\multicolumn{2}{c}{\textbf{Total input}} & 
\multicolumn{4}{c}{\textbf{Total output}} &
\makecell{\textbf{SUM}} \\
\cmidrule(lr){2-3} \cmidrule(lr){4-7}
 & 
\makecell{\textbf{$Token_{in}$}} & 
\makecell{\textbf{$Cost_{in}$}} & 
\makecell{\textbf{$Token_{rea}$}} & 
\makecell{\textbf{$Cost_{rea}$}} & 
\makecell{\textbf{$Token_{out}$}} & 
\makecell{\textbf{$Cost_{out}$}} &
 \\
\midrule
GPT-5 & 82622 & 0.103 & 204237 & 2.042 & 4016 & 0.040 & 2.186 \\
o3 & 88152 & 0.176 & 159770 & 1.278 & 5927 & 0.047 & 1.502 \\
Gemini 2.5 Pro & 30600 & 0.038 & 403024 & 4.030 & 61533 & 0.615 & 4.684 \\
GPT-5 mini & 148923 & 0.037 & 112128 & 0.224 & 3949 & 0.008 & 0.269 \\
o4 mini & 211157 & 0.232 & 168288 & 0.741 & 7763 & 0.034 & 1.007 \\
Gemini 2.5 Flash & 30600 & 0.009 & 264592 & 0.662 & 71756 & 0.179 & 0.850 \\
\makecell[l]{Gemini 2.5 Flash\\(no reasoning)} & 30600 & 0.009 & -- & -- & 122267 & 0.306 & 0.315 \\
GPT-5 nano & 184807 & 0.009 & 278515 & 0.111 & 2880 & 0.001 & 0.122 \\
Claude Opus 4 & 138554 & 2.078 & -- & -- & 32812 & 2.461 & 4.539 \\
GPT-4.1 mini & 159056 & 0.064 & -- & -- & 19759 & 0.032 & 0.095 \\
Claude Sonnet 4 & 138554 & 0.416 & -- & -- & 37799 & 0.567 & 0.983 \\
GPT-4.1 & 99314 & 0.199 & -- & -- & 14777 & 0.118 & 0.317 \\
Gemini 2.0 Flash & 231738 & 0.023 & -- & -- & 9837 & 0.004 & 0.027 \\
GPT-4o & 99314 & 0.248 & -- & -- & 15585 & 0.156 & 0.404 \\
Gemma 3-27b & 30600 & 0 & -- & -- & -- & 0 & 0 \\
Claude Haiku 3.5 & 138702 & 0.111 & -- & -- & 13324 & 0.053 & 0.164 \\
GPT-4.1 nano & 299810 & 0.030 & -- & -- & 12009 & 0.005 & 0.035 \\
Gemma 3-4b & 30600 & 0 & -- & -- & -- & 0 & 0 \\
\botrule
\end{tabular*}
\end{table*}

Overall, these results suggest that model selection for educational applications should be guided by empirical performance data rather than assumptions based on provider reputation or listed per-token prices. Lower-cost models may, in some cases, offer equal, comparable, or even superior performance on physics conceptual tasks involving visual representations.




\section{Discussion, limitations and future work}\label{Discussion}

The analysis of performance shows that the best-performing MLLMs are already outperforming post-instruction university student averages on the four tested concept inventories. Some of them are coming close to expert-like performance on some of the inventories. The top-tier models from OpenAI and Google are now exceeding 80\% and 75\% accuracy, respectively, on our benchmark, suggesting real potential for educational deployment. However, the non-uniform performance across the four underlying concept inventories suggests that even the best models still struggle with many items. While the tested models collectively performed quite well on BEMA and TUG-K, their scores dropped significantly on QMVI and, most notably, FTGOT. These findings suggest a need for future research focused on analyzing how specific visual formats and task types impact model performance, as well as the need for qualitative studies examining the reasoning behind MLLM-selected answers.

The cost analysis reveals that performance does not scale linearly with costs. While some of the most capable models remain relatively cost-effective (e.g., GPT-5 mini, o4 mini and o3 mini), others—like Claude Opus 4—underperformed despite high costs. This suggests that institutions cannot rely solely on pricing tiers (cost per token) or provider reputation as proxies for quality across different educational settings. Conversely, certain mid-range or low-cost models offer a compelling balance of performance and affordability. In particular, GPT-5 mini stands out as a well-performing and low-cost model. GPT-5 nano and Gemini 2.5 Flash with ``disabled reasoning" also achieved over 60\% average accuracy at a fraction of the cost of more expensive models. Our results thus indicate that more affordable models can reach performance levels that are close to those of the best performing models. This has important implications for deploying  AI  in schools or universities operating under financial constraints. However, freely available or open-weight models (e.g., Gemma 3 series) currently perform well below acceptable thresholds for educational use on physics tasks containing images. These models may be interesting for other roles, but are not yet suitable for student-facing educational applications or assistance with assessment or grading when physics images are involved.

While this study offers a broad and comparative view of MLLM performance on conceptual physics tasks involving images, it also has several limitations.

First, our analysis focused exclusively on multiple-choice items from well-established concept inventories. While this design allows for standardization and comparability, it does not capture how MLLMs perform on more open-ended or ill-structured physics tasks, including derivations, written explanations, or lab-based data analysis that better represent authentic assessment practices in physics classrooms. Future research should expand to include these formats, which are common in authentic classroom and assessment settings.

Second, our evaluation was quantitative in nature. We scored responses based solely on the selected answer choices, without analyzing the correctness or coherence of the generated output text. This leaves open the possibility that models selected correct answers for the wrong reasons—or, conversely, generated valid reasoning but selected incorrect options~\cite{tugk, bema}. A qualitative investigation of model-generated explanations would be a valuable next step to better understand reasoning quality and error patterns. Some models (e.g. OpenAI's reasoning models) offer so-called ``reasoning summaries" as an optional part of their output~\cite{openaiguide}. A reasoning summary is a condensed recap of the content of the reasoning that the model did in the background, as it employed the internal CoT process hidden from the user, to solve the task at hand. Future research could qualitatively investigate these ``reasoning summaries" to get better insights into the strengths and weaknesses of these models. Such analyses could also help determine which types of visual tasks pose the greatest challenges, and why certain inventories consistently produce lower scores even for otherwise high-performing models. For example, follow-up studies could attempt to disentangle model strengths and weaknesses in spatial, symbolic, and graphical reasoning. This can be done by collecting new data, or re-analyzing the data collected in this project~\cite{zenodo}.

Third, all the models were evaluated using a static, minimal prompt. We intentionally avoided further prompt engineering to reflect default user scenarios and preserve comparability. However, this likely underestimates the full potential of certain models, particularly those responsive to CoT prompting or domain-specific scaffolding. Future studies could explore the impact of tailored prompts. It is important to note, though, that OpenAI explicitly advises against CoT prompting for its reasoning models (e.g., GPT-5 series, o3, o4-mini), stating that the approach is likely to be ineffective or can even worsen the models' performance~\cite{openaibestpractice}. For ``non-reasoning" models, on the other hand, performance might benefit from more specialised CoT prompting. However, this would likely also lead to an increased number of generated tokens and consequently a higher cost. Systematic evaluation is needed to determine if CoT prompting with non-reasoning models is economically preferable to running reasoning models. Another consideration that can be relevant from an educational perspective is the transparency of the model output. Using models that do not hide part of their output from the user can be preferred, especially when the focus lies on the process of getting to a solution, not just the final answer.  In this respect, educators need to experiment with different models and prompting approaches to achieve the desired behaviour for their praticular use case (e.g., tutoring, grading, feedback).

Finally, this study represents a snapshot in time. The capabilities, pricing, and availability of MLLMs are evolving rapidly, and new model releases or fine-tuned educational variants may soon outperform those tested here. Maintaining up-to-date benchmarks and developing open testing protocols will continue to be important for tracking progress and supporting informed decision-making.

By addressing these limitations and building on our current findings, future research can deepen our understanding of how and when MLLMs can meaningfully contribute to physics education—and where caution, adaptation, or complementary approaches remain necessary.




\section{Conclusion}\label{Conclusion}

This study provides a comparative evaluation of a selection of publicly available MLLMs on conceptual physics tasks requiring visual interpretation. By benchmarking both performance and cost across multiple concept inventories, we highlight critical differences among models that are not apparent from pricing or provider claims alone.

Our findings suggest that some MLLMs now approach expert-level accuracy on certain physics concept inventories in domains such as kinematics and electromagnetism. However, performance drops significantly on tasks involving complex spatial or abstract reasoning, particularly in geometrical optics. On the other hand, the analysis shows that high performance does not necessarily come with high cost. Several models offer favorable cost–performance ratios, making them viable options for educational deployment, including in resource-constrained settings. Conversely, some of the most expensive models underperformed, suggesting that informed model selection is crucial.

As MLLMs continue to improve and gain traction in educational applications, physics educators and institutions must continuously and critically evaluate them, and carefully consider the balance of capability, cost, and context-specific needs. This study offers an example of an evaluation that can help physics educators in this process.




\bibliography{sn-bibliography}

\end{document}